# Can compressed sensing beat the Nyquist sampling rate?

**Leonid P. Yaroslavsky,**

*Dept. of Physical Electronics, School of Electrical Engineering, Tel Aviv University, Tel Aviv 699789, Israel*

Data saving capability of "Compressed sensing (sampling)" in signal discretization is disputed and found to be far below the theoretical upper bound defined by the signal sparsity. On a simple and intuitive example, it is demonstrated that, in a realistic scenario for signals that are believed to be sparse, one can achieve a substantially larger saving than compressing sensing can. It is also shown that frequent assertions in the literature that "Compressed sensing" can beat the Nyquist sampling approach are misleading substitution of terms and are rooted in misinterpretation of the sampling theory.

This letter is motivated by recent OPN publications[1,2] that advertise wide use in optical sensing of "Compressed sensing" (CS), a new method of digital image formation that has obtained considerable attention after publications[3-7]. This attention is driven by such assertions in numerous publications as "CS theory asserts that one can recover certain signals and images from far fewer samples or measurements than traditional methods use"[6], "Beating Nyquist with Light"[1,8,9] and such. For those, who are familiar with sampling theory and know that the Nyquist rate can't be beaten, these assertions sound questionable. Is that true that "Compressed sensing guarantees on reconstruction quality even when sampling is very far from Nyquist sampling", and, if yes, how much one can win in terms of reducing the sampling rate? In what follows we attempt to answer these questions.

Compressed sensing assumes signal approximation by their "sparse" copies, i.e. signals whose spectrum in a selected transform has only few non-zero components. According to this approach, one should specify the total number $N$ of signal samples required for its discrete representation and the number $M<N$ of certain measurements to be done in order to obtain, by means of signal $L1$ norm minimization in a selected transform domain, signal sparse discrete representation with $N$ samples and certain $K<M$ non-zero spectral components. The ratio $CSDRF=N/M$ (CS Dimensionality Reduction Factor) is the degree of the achieved signal dimensionality reduction. The ratio $SPRS=K/N$ of the number $K$ of non-zero transform coefficients to the total number $N$ of signal samples is called the signal sparsity.

The theoretical upper bound of signal dimensionality reduction that can be achieved by means of signal "sparse" approximation can be evaluated using the Discrete Sampling Theorem, according to which if a signal of $N$ samples has only $K$ non-zero components of its spectrum in a certain transform, the minimal number of its samples sufficient for its perfect restoration is $K$[13]. For such transforms as Discrete Fourier and Discrete Cosine Transform (DCT), positions of these $K$ samples can be arbitrary.

*Address all correspondence to: Leonid. P. Yaroslavsky, E-mail: yaro@eng.tau.ac.il

Therefore, the inverse of the signal sparsity $SPRS=K/N$ is the theoretical upper bound of signal dimensionality reduction achievable by means of signal sparse approximation. This relationship is plotted in Fig. 1 (solid line).

The upper bound of signal dimensionality reduction $CSDRF$ of signals with sparsity SSP achievable by CS can be evaluated from the relationship: $1/CSDRF>C(2SSP\log CSDRF)$ provided in Ref. **Error! Reference source not found.**], where $C>1$ and $K<M<<N$. This relationship is used to plot in Figure 1 two curves, $SSP<1/(2CSDRF\log CSDRF+1)$ (dash-dot line) and $SSP<1/(3.5CSDRF\log CSDRF+1)$ (dot line) ones that asymptotically, for $K<M<<N$, tend to it for values 1 and 1.75 of the multiplier $C$, correspondingly, and, from the other side, satisfy the natural requirement that, for $K=N$ there must be $M=N$. The first curve can be considered as a theoretical upper bound of the signal dimensionality reduction capability of CS. The second curve better fits experimental data, shown by diamonds in Figure 1, on experimental values of CS dimensionality reduction factors found in the literature and listed, along with the corresponding sources, in **Error! Reference source not found.**. One can regard it as an experimental upper bound.

As one can see, CS requires a substantially redundant number of data with respect to the theoretical bound defined by the inverse to signal sparsity. One can numerically evaluate the degree of the redundancy from plots in Fig. 2 of ratios of this bound and CS theoretical (dash-dot line) and experimental (dot line) bounds versus signal spectrum sparsity $K/N$.

This sampling redundancy of CS is the price one should pay for the uncertainty regarding indices of signal non-zero spectral components: CS approach assumes the belief that signals can be approximated by their "sparse" copies but does not assume any specification of positions of signal non-zero spectral components.

However this total uncertainty is a too pessimistic scenario. If one believes that an image can be satisfactorily approximated by its copy sparse in a certain transform, as a rule, one knows, at least roughly, from energy compaction properties of the transform, where image nonzero transform coefficients are expected to be concentrated. Even such a vague knowledge can greatly help.

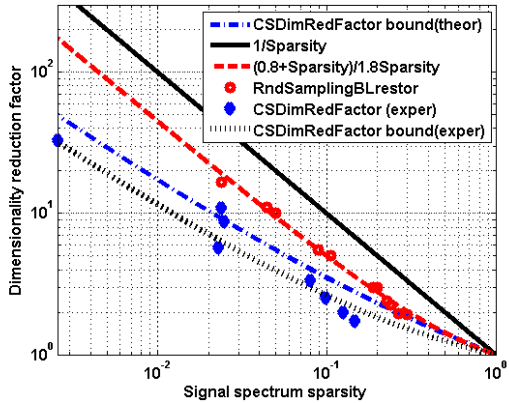

Fig. 1 Signal dimensionality reduction factor versus spectrum sparsity: theoretical upper bound (solid), theoretical (dash-dot) and experimental (dot) upper bounds for Compressive sensing and its estimate for random sampling and band-limited reconstruction (dash)

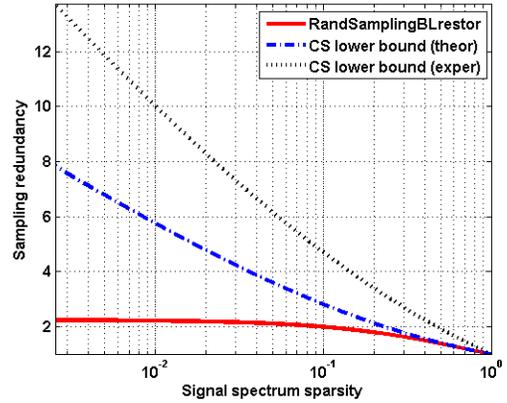

Fig. 2. Sampling redundancy of compressed sensing and of random sampling and band-limited approximation versus image sparsity

For instance, for an overwhelming number of real images, appropriate transforms such as DCT compact image energy into the lower frequency part of spectral components. It is this property that is put in the base of transform coefficient zonal quantization tables in transform image coding such as JPEG. Therefore, one can, in addition to specifying the number N of desired images samples and the number M of samples to be taken, which is anyway required by the CS approach, make a natural assumption that image spectral components important for image reconstruction are concentrated within, say, a circular shape that encompasses M spectral components with the lowest indices. With this assumption, one can either sample the image in a regular way with a sampling rate defined by dimensions of a square that circumscribes this shape or, which is more efficient in terms of reducing the required the number of samples, reconstruct an image sparse approximation from a set of M samples taken, in the case of sparsity of DCT or DFT spectra, in randomly chosen positions. For the reconstruction, an iterative Gershberg-Papoulis type algorithm can be employed[13]. This option (let's call it "Random Sampling and Band Limited Reconstruction", or RSBLR), is illustrated in Figure 3 on an example of a test image "Ango" from a set of 11 test images listed in Table 2.

The values of image dimensionality reduction N/M found in experiments with these images are plotted in Fig. 1 (bold circles) along with a curve (0.8+Sparsity)/(1.8*Sparsity) that fits them sufficiently well (dash line). Solid line in Fig. 2 represents an estimate of the sampling redundancy of RSBLR obtained as a ratio of the sparse approximation dimensionality redundancy theoretical upper bound (solid line in Fig. 1) to the fitting curve (0.8+*Sparsity*)/(1.8**Sparsity*). As sparsity of images can be specified only for a certain quality of image sparse approximation, it was opted to use the quality of standard JPEG image compression as an acceptable quality of image approximation, i.e. sparsity of test image spectra (second column of Table 2) was evaluated in these experiments as a fraction of image DCT most intensive spectral coefficients, which are sufficient for image reconstruction with the same RMS reconstruction error as that for the standard JPEG compression of the corresponding test image (fifth column).

As one can see from Fig. 2, Random Sampling and Band Limited Reconstruction substantially outperforms CS in its signal dimensionality reduction efficiency practically in the entire range of values of possible image sparsities.

Note that in less common cases, when one can believe that image nonzero spectral coefficients are concentrated not only in low frequencies but in few disjoint areas within spectral domain, this method can be employed as well. Again, even vague knowledge regarding positions of image nonzero spectral components, which is associated with this belief, can give a substantial saving in the number of required image samples compared to the totally "blind" restoration using CS approach.

Table 1. Experimental data on image dimensionality reduction achieved by using compressed sensing methods

| Source | Spectrum sparsity | Dimensionality reduction factor |
|---|---|---|
| Ref. 6, Fig. 3 | 0.125 | 2 |
| Ref. 6, Fig. 1 | 0.0238 | 10.92 |
| Ref. 1, p. 48 | 0.00045 | 33 |
| Ref. 10 | 0.0238 | 5.67 |
| Ref. 10 | 0.0238 | 8.56 |
| Ref. 11 | 0.0992 | 2.52 |
| Ref. 12, p.1 | 0.08 | 3.33 |
| Ref. 12, p.2 | 0.146 | 1.73 |

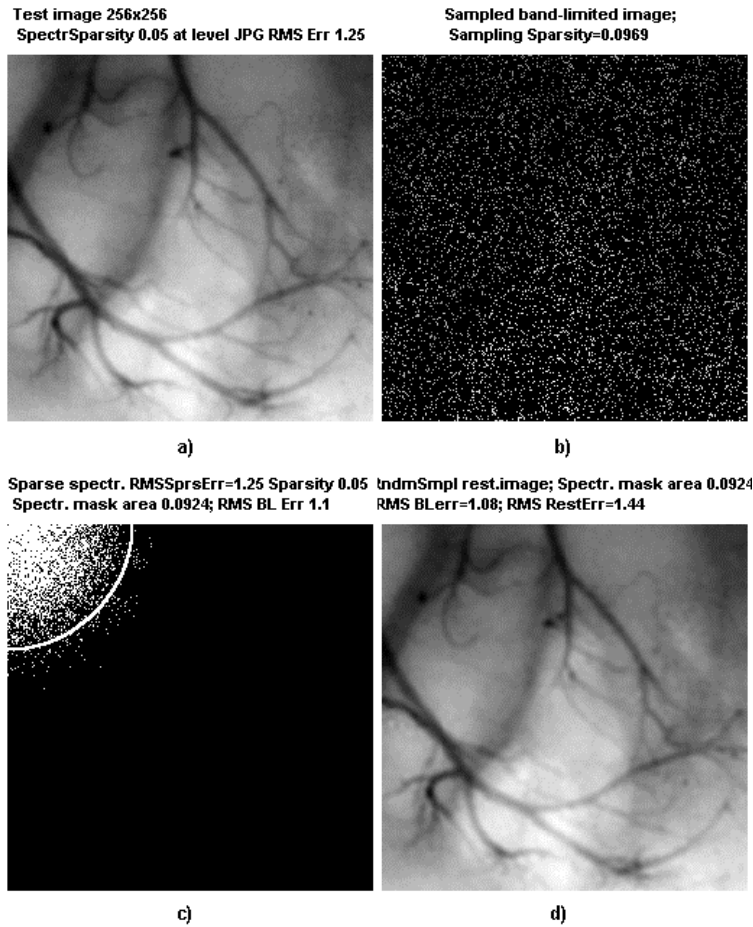

Fig. 3. (a): Test image of 256x256 pixels, (b): its 6141 samples taken in randomly selected positions, (c): a map (white dots) of 3277 most intensive components of its DCT spectrum, which reconstruct the image with the same RMS error of 1.25 gray levels as that of its JPEG reconstruction error, and the border (white line) of the low pass filter that encompasses 6029 DCT spectral components with the lowest indices; (d): the result of the band-limited reconstruction with RMS error of 1.46 gray levels from the sparse random samples.

Consider now assertions that CS can "beat the Nyquist sampling approach"1. CS technique can, in principle, restore signals with few spectral components within the base-band defined by the component of the highest frequency from their samples taken with a rate lower than twice this frequency. This, however, certainly does not mean at all that it beats the Nyquist sampling. The reason is very simple: twice the component highest frequency is not the minimal sampling rate for such signals. According to the sampling theory, for sampling signals that contain few spectral components, the minimal sampling rate is defined by the total area occupied by signal spectral components in transform domain. Optimal sampling of such signals is sub-band sampling, which requires signal sinusoidal modulation-demodulation in order to shift signal high frequency sub-bands to a low frequency band before sampling and then to shift them back for signal reconstruction. This fundamental result of the sampling theory dates back to 1950-th - 1960-th and is addressed in many publications and textbooks. Among the most recent ones are Refs. 14 and 15.

CS replaces signal sinusoidal modulation-demodulation by signal blind modulation-demodulation using pseudo-random masks, but pays quite a high price of substantial redundancy in the required number of samples. For instance, in the experiment of sampling and reconstruction of a high frequency sinusoidal signal presented in Ref. 1, the required redundancy (the ratio M/K, in denotations of the paper) is 1/0.015≈67 times. Note that no analysis concerning the accuracy of signal high frequency sinusoidal components restoration and possible aliasing artifacts is provided in that publication, as well as in others similar.

To conclude, the theoretical estimates and experimental data presented above show that assertions that CS methods enable large reduction in the sampling costs and surpass the traditional limits of sampling theory are quite exaggerated, misleading and grounded in misinterpretation of the sampling theory.

Table 2. Experimental data on image dimensionality reduction achieved by using random sampling and band limited reconstruction (RSBLR) method for a set of test images

| Test image | Spectrum sparsity $K/N$ | RSBLR Dimensionality reduction factor $N/M$ | RMS error: RSBLR reconstruction | RMS error: JPEG reconstruction |
|---|---|---|---|---|
| Mamm | 0.044 | 11 | 1.58 | 1.48 |
| Ango | 0.05 | 10.5 | 1.36 | 1.25 |
| Test4CS | 0.89 | 5.55 | 2.15 | 1.62 |
| Moon | 0.105 | 5 | 2.55 | 2.5 |
| Lena512 | 0.19 | 3 | 3.3 | 3.9 |
| Aerial photo | 0.2 | 3 | 4.76 | 4.5 |
| Man | 0.227 | 2.38 | 4.12 | 4 |
| ManSprse [6] | 0.024 | 16.7 | 5.6 | - |
| Multiphoton | 0.238 | 2.26 | 4.73 | 4.89 |
| Barbara512 | 0.265 | 1.61 | 4.15 | 3.91 |
| Westconcord | 0.301 | 1.92 | 7.48 | 7.21 |